# Detailed kinetic models for the low-temperature auto-ignition of gasoline surrogates

R. Bounaceur, P.A. Glaude, R. Fournet, V. Warth, F. Battin-Leclerc[*]

Département de Chimie Physique des Réactions, CNRS-Nancy University, FRANCE

**Abstract**
In the context of the search for gasoline surrogates for kinetic modeling purpose, this paper describes a new model for the low-temperature auto-ignition of n-heptane/iso-octane/hexene/toluene blends for the different linear isomers of hexene. The model simulates satisfactory experimental results obtained in a rapid compression machine for temperatures ranging from 650 to 850 K in the case of binary and ternary mixtures including iso-octane, 1-hexene and toluene. Predictive simulations have also been performed for the autoignition of n-heptane/iso-octane/hexene/toluene quaternary mixtures: the predicted reactivity is close to that of pure iso-octane with a retarding effect when going from 1- to 3-alkene.

**Introduction**

While gasoline and diesel fuel have a "near-continuous spectrum" of hydrocarbon constituents, surrogates composed of a limited number of components have to be defined in order to develop detailed kinetic models. The following strategy to develop them has been proposed [1]:
"1. Feasibility. Candidates in the formula must have known detailed kinetic mechanisms.
2. Simplicity. Mainly limited for computational capabilities to normal paraffins with less than 12 carbon atoms, monocyclic paraffins with less than 8 carbon atoms, and simple aromatics such as benzene, alkyl-benzenes and naphthalene.
3. Similarity. The surrogate is required to match practical fuels on both physical and chemical properties: (i) volatility (boiling range and flash point), (ii) sooting tendency (smoking point and luminous number), (iii) combustion property (heat of combustion, flammability limits and laminar premixed mass burning rate).
4. Cost and availability."

The hydrocarbon constituents of gasoline contain from 4 to 10 atoms of carbon and can be divided in five main families, namely linear alkanes (n-paraffins, for about 10% mass in an European gasoline), branched alkanes (iso-paraffins, for about 30%), cyclic alkanes (naphtenes, for about 3%), alkenes (olefins, for about 20 %) and aromatic compounds (for about 35%) [2]. Since normal-heptane and iso-octane are primary reference fuels (PRF) for octane rating in spark-ignited internal combustion engines, the n-heptane/iso-octane blend (PRF mixture) has long been the most commonly proposed surrogate to reproduce low-temperature oxidation of gasoline (e.g. [3]). More recently, models have been proposed for a PRF/toluene blend [4,5], as well as for a n-heptane/iso-octane/1-pentene/toluene/methylcyclohexane mixture [6].

The low-temperature auto-ignition of a ternary iso-octane/1-hexene/toluene blend and of the related binary mixtures has also been recently investigated: while satisfactory modeling was obtained for the iso-octane/1-hexene mixture, the agreement deteriorated for the other blends [7-8]. No experimental data are available for hydrocarbons mixtures including an alkene in which the double bond is not the first one.

The purpose of the present paper is to describe new models for the low-temperature auto-ignition characteristics of n-heptane/iso-octane/hexene/toluene blends for the different linear isomers of hexene, to validate that related to 1-hexene using the existing experimental data and finally to use them to study the influence of the position of the double bond on the reactivity of n-heptane/iso-octane/hexene/toluene mixtures representative of gasoline.

**Description of the model**

A recent mechanism for toluene from our team [9] has been added to models generated using the EXGAS system for n-heptane/iso-octane/hexene mixtures. The mechanisms generated by EXGAS for the oxidation of n-heptane and iso-octane taken individually or in mixtures have been previously tested [10]. Models for the oxidation of the different linear isomers of hexene have been recently developed by Bounaceur et al. [11]. All these models have been validated in the same temperature range as the present study. In the case of 1-hexene, the full mechanism involves 1257 species and includes 5803 reactions, while a mechanism for the oxidation of n-heptane/iso-octane/toluene would only includes 2575 reactions.

It is worth noting that the presence of a double bond in alkene molecules involves an important complexity of the chemistry of low temperature oxidation. The radicals directly deriving from the reactant are no longer of a single type, as alkyl radicals from alkanes, but of at least three types, alkylic and allylic alkenyl radicals being obtained by H-abstraction and hydroxyalkyl radicals being obtained by addition of •OH radicals to the double bond. This explains the scarcity of models related to alkenes, representative of the compounds present in gasoline, even if these unsaturated compounds are the major products of the oxidation of alkanes.

*General features of the EXGAS system for the modeling of the oxidation of alkanes and alkenes*

As several papers already described the way EXGAS generates detailed kinetic models for the oxidation of alkanes [10, 12] and alkenes [11, 13, 14], only a summary of its main features is given here. The system provides reaction mechanisms made of three parts:

♦ A $C_0$-$C_2$ reaction base, including all the reactions involving radicals or molecules containing less than three carbon atoms [15].

♦ A comprehensive primary mechanism including all the reactions of the molecular reactants, the initial organic compounds and oxygen, and of the derived free radicals.

♦ A lumped secondary mechanism, containing the reactions consuming the molecular products of the primary mechanism which do not react in the reaction base.

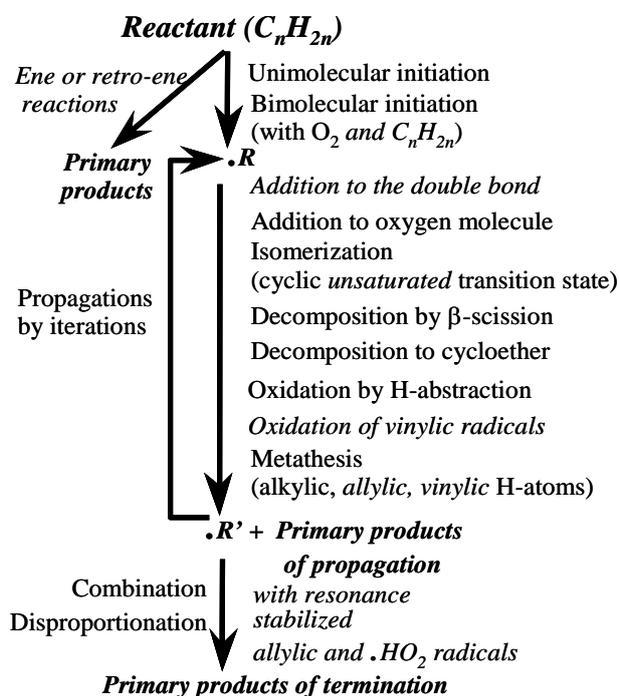

*Figure 1: Algorithm of generation for the primary mechanism of the oxidation of alkanes and alkenes. The reactions in italics are those specific to alkenes.*

Figure 1 summarizes the different types of generic reactions, which are taken into account in the primary mechanism of the oxidation of alkanes and alkenes, and the structure of the algorithm, which is used for the generation and which ensures the comprehensiveness of the mechanisms. This figure also shows which reactions are considered for both types of reactants and which are only taken into account in the case of alkenes. Previous papers [11, 13, 14] describe in details these generic reactions and the related kinetic parameters.

To ensure the primary mechanism to be comprehensive, all the isomers of reactants, primary free radicals and molecules are considered. Therefore, during the generation of the primary mechanism, both EXGAS and the connected softwares use the detailed chemical formulae of all these molecules and free radicals, reactants and products. In order to keep a manageable size, the lumped secondary mechanisms involved lumped reactants: the molecules formed in the primary mechanism, with the same molecular formula and the same functional groups, are lumped into one unique species without distinction between the different isomers. Up to now, these secondary mechanisms included global reactions which mainly produced, in the smallest number of steps, molecules or radicals, the reactions of which are included in the $C_0$-$C_2$ reactions base.

Thermochemical data for molecules or radicals are automatically calculated and stored as 14 polynomial coefficients, according to the CHEMKIN formalism [16]. These data are computed using the THERGAS software [17], based on group or bond additivity and thermochemical kinetics methods proposed by S. Benson (1976) [18].

Kinetic data are either calculated using the KINGAS software [12] based on the thermochemical kinetics methods proposed by S. Benson (1976) [18] or estimated through quantitative structure-reactivity relationships [11, 13, 14].

*A model for the oxidation of toluene*
The model for the oxidation of toluene includes the following sub-mechanisms [11]:

♦ A primary mechanism including reactions of toluene containing 193 reactions and including the reactions of toluene and of benzyl, tolyl (methylphenyl), peroxybenzyl, alcoxybenzyl and cresoxy free radicals.

♦ A secondary mechanism involving the reactions of benzaldehyde, benzyl hydroperoxyde, cresol, benzylalcohol, ethylbenzene, styrene and bibenzyl.

♦ A mechanism for the oxidation of benzene [19] including the reactions of benzene and of cyclohexadienyl, phenyl, phenylperoxy, phenoxy, hydroxyphenoxy, cyclopentadienyl, cyclopentadienoxy and hydroxycyclopentadienyl free radicals, as well as the reactions of ortho-benzoquinone, phenol, cyclopentadiene, cyclopentadienone and vinylketene.
- ♦ A mechanism for the oxidation of unsaturated $C_0$-$C_4$ species, which contains reactions involving •$C_3H_2$, •$C_3H_3$, $C_3H_4$ (allene and propyne), •$C_3H_5$ (three isomers), $C_3H_6$, $C_4H_2$, •$C_4H_3$ (2 isomers), $C_4H_4$, •$C_4H_5$ (5 isomers), $C_4H_6$ (1,3-butadiene, 1,2-butadiene, methyl-cyclopropene, 1-butyne and 2-butyne).

This mechanism for the oxidation of toluene has been validated in previous work using experimental results obtained in jet stirred and plug flow reactors and in shock tubes.

*Crossed reactions*

The software EXGAS is designed to consider systematically the crossed reactions: each time a radical is created, it is submitted to all possible generic propagations without considering the reactant that first yielded it. The possible crossed reactions are of three types:

- ♦ *Crossed reactions between species deriving from alkanes and hexenes.*

The metatheses involving the abstraction of a hydrogen atom from a hexene molecule by heptyl and iso-octyl radicals, as well as the combinations between allylic hexenyl and alkyl radicals have been considered. These reactions are shown hereafter in the case of 1-hexene:

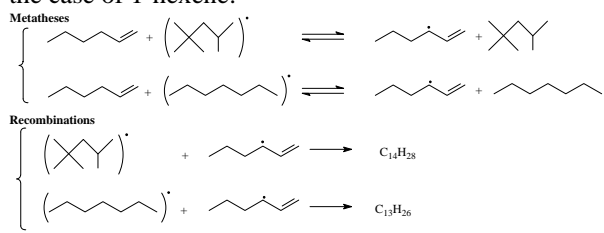

- ♦ *Crossed reactions between species deriving from alkanes and toluene.*

The metatheses involving the abstraction of a hydrogen atom from n-heptane or iso-octane by benzyl radicals, as well as the combinations between benzyl and alkyl radicals have been written, as shown below:

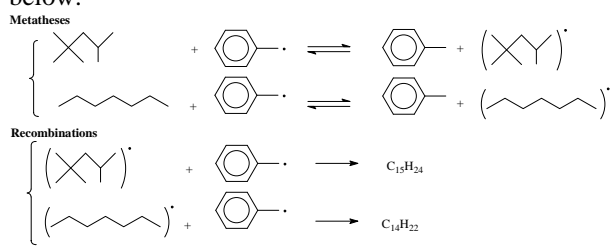

- ♦ *Crossed reactions between species deriving from toluene and hexenes.*

The metatheses involving the abstraction of a benzylic hydrogen atom from toluene by allylic radicals have been taken into account.

**Results and discussion**

Simulations have been performed using the SENKIN software of CHEMKIN II [16]. Validations have been made for the ternary iso-octane/1-hexene/toluene blend and for the three different related binary mixtures using for comparison the experimental results obtained in a rapid compression machine. Predictive simulations have been run under similar conditions for n-heptane/iso-octane/hexene/ toluene blends including the different linear isomers of hexene. In all the figures, symbols correspond to experimental results and lines to simulations.

*Validations for binary and ternary mixtures containing iso-octane, 1-hexene or toluene*

Vanhove et al. [20] have measured cool flame and auto-ignition delay times in a rapid compression machine for iso-octane/1-hexene/toluene/$O_2$/Ar/ $N_2$/$CO_2$ mixtures for temperatures ($T_c$) after compression ranging from 650 to 850 K. $T_c$ is calculated at the end of the compression based on an adiabatic core gas model and can be considered as the maximum temperature reached in the combustion chamber. The three binary mixtures and the ternary one have been investigated. For each mixture, different initial pressures have been tested leading to different pressures after compression ($P_c$).

In the case of the binary mixtures, the agreement obtained between our simulations and the experimental results is satisfactory both for cool flame and ignition delay times for all studied pressure ranges, as illustrated in Figure 2 in the case of the 11-15 bar range. Simulations reproduce well the difference of reactivity between the three mixtures.

The worst agreement between simulations and experimental results is obtained for the iso-octane/toluene mixture, for which ignition delay times experimentally and numerically recorded are much larger than for the two other blends and for which any cool flame was neither experimentally observed nor simulated. The reactivity of the iso-octane/1-hexene mixture is larger than that of the toluene/1-hexene blend and a negative temperature coefficient (NTC) zone in only observed, as well as predicted by the model, in the first case. As shown by Vanhove et al. [7], iso-octane has no retarding effect on the autoignition of 1-hexene below 750 K, since the reactivity of the iso-octane/1-hexene mixture is close to that of pure 1-hexene. That is not the case of toluene which considerably slows down the low-temperature auto-ignition of 1-hexene, as well as that of iso-octane.

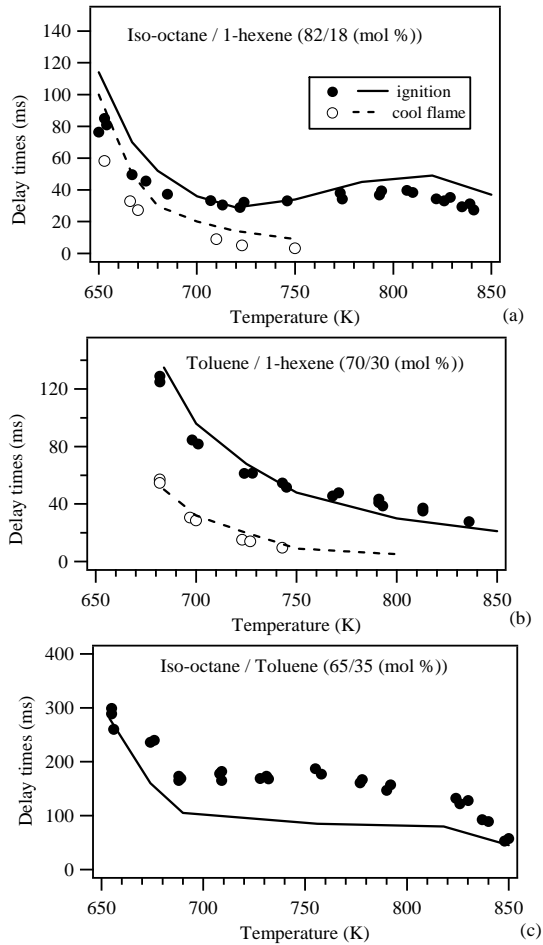

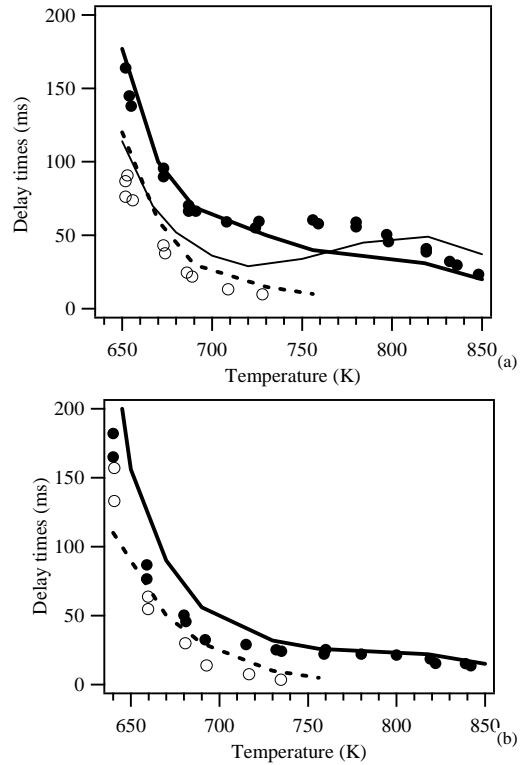

*Figure 2: Cool flame and ignition delay times for the three binary mixtures in a rapid compression machine ($P_c$ between 11 and 15 bar, stoichiometric mixtures in air, experimental results from Vanhove et al. [20]).*

*Figure 3: Cool flame and ignition delay times for a (47/35/18 % mol) iso-octane/tolene/1-hexene ternary mixture in a rapid compression machine for (a) $P_c$ between 11 and 15 bar and (b) $P_c$ between 15 and 20 bar (stoichiometric mixtures in air, experimental results from Vanhove et al., 2006a). The thin line in (a) is the simulation for the (82/18 % mol) iso-octane/1-hexene binary mixture under the same conditions as shown in fig 2a.*

As shown in figure 3, the agreement is also satisfactory in the case of a ternary mixtures containing 47% (mol) iso-octane, 35% toluene and 18% 1-hexene for pressures after compression ranging from 11 to 15 bar and from 15 to 20 bar.

Between 11 and 15 bar, the ternary mixture is much less reactive than the iso-octane/1-hexene binary mixture at the lowest temperatures (below 750 K). The presence of toluene also makes the NTC zone disappear in both experimental results and simulations. The decrease of the cool flame and ignition delay times when pressure increases is well reproduced.

*Predictive simulations for n-heptane/iso-octane/ hexene/toluene blends including the different linear isomers of hexene*

Figure 4 presents the results of a simulation for the auto-ignition of three quaternary mixtures containing 8% (mol) n-heptane, 44% iso-octane, 30% toluene and 18% hexene for each linear isomer of hexene. This composition has been chosen to be close of that of an actual gasoline. This figure exhibits a close reactivity for these three mixtures, with differences in ignition delay times being never larger than a factor 1.5. However, simulations show a difference of reactivity according to the included isomer of hexene. While 2- and 3-hexene have close octane numbers (RON is 76.4, 92.7 and 94, for 1-hexene, 2-hexene and 3-hexene, respectively, [2]), the reactivity of the mixture containing 2-hexene is much more similar to that including 1-hexene than to that containing 3-hexene.

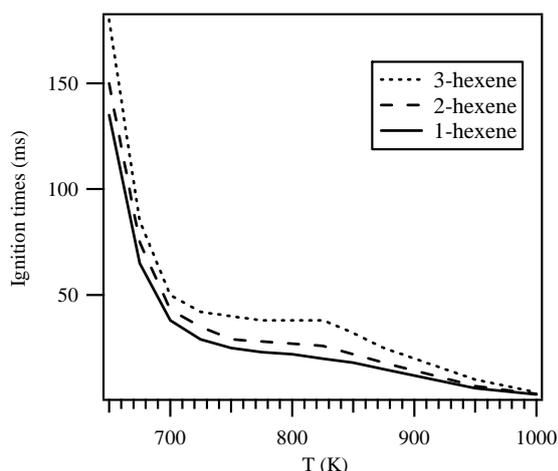

*Figure 4: Ignition delay times for three (8/44/30/18 % mol) n-heptane/iso-octane/toluene/hexane quaternary mixtures for each linear isomer of hexene (P = 15 bar, stoichiometric mixtures in air).*

Figure 5 display ignition delay times of the quaternary mixture including 1-hexene compared to those of pure n-heptane, iso-octane and 1-hexene at a pressure of 15 bar. Toluene does not ignite under these conditions. This figure shows that, under the same conditions, the behavior of the quaternary mixture including 1-hexene is very close to that of iso-octane at the lowest temperatures (i.e. below 720 K), but that no NTC zone is predicted. The retarding effect of the presence of toluene counteracts the promoting influence induced by 1-hexene and n-heptane which are very reactive fuels.

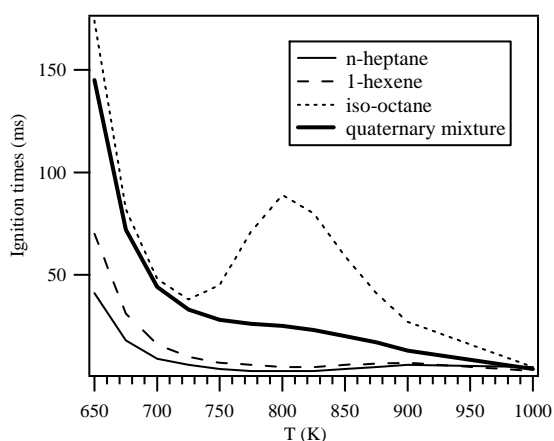

*Figure 5: Comparison between the ignition delay times of the quaternary mixture including 1-hexene and those of pure n-heptane, iso-octane and 1-hexene (P = 15 bar, stoichiometric mixtures in air).*

**Conclusion**

This paper presents a new model for the low-temperature auto-ignition of n-heptane/iso-octane/hexene/toluene blends for the different linear isomers of hexene. These models, which include up to 5800 reactions, allow satisfactory simulations compared to experimental ignition delay times measured in a rapid compression machine.

The use of such large models can only be fully justified if not only ignition delay times are well modeled, but also the products distribution. But that can only be achieved when more experimental data on the formation of products are available.

**Acknowledgement**
This work has been supported by Institut Français du Pétrole.